\documentclass[conference]{IEEEtran}
\IEEEoverridecommandlockouts
\usepackage{cite}
\usepackage{amsmath,amssymb,amsfonts,bm}
\usepackage{theorem}
\usepackage{graphicx}
\usepackage{textcomp}
\usepackage{url}
\usepackage{xcolor}
\usepackage[nolist,nohyperlinks]{acronym}
\usepackage{enumitem}
\usepackage[ruled,vlined,titlenumbered,linesnumbered]{algorithm2e}
\def\BibTeX{{\rm B\kern-.05em{\sc i\kern-.025em b}\kern-.08em
    T\kern-.1667em\lower.7ex\hbox{E}\kern-.125emX}}

\newtheorem{theorem}{Theorem}
\newtheorem{definition}{Definition}

\newcommand{\Fqm}{\ensuremath{\mathbb F_{q^m}}}
\newcommand{\Fqn}{\ensuremath{\mathbb F_{q^n}}}

\newcommand{\Fq}{\ensuremath{\mathbb F_{q}}}

\newcommand{\F}{\ensuremath{\mathbb F}}

\DeclareMathOperator{\defi}{def}
\newcommand{\defeq}{\overset{\defi}{=}}

\renewcommand{\mod}{\; \textnormal{ mod } \;}

\DeclareMathOperator{\rk}{rk}

\renewcommand{\vec}[1]{\ensuremath{\bm{#1}}}
\newcommand{\trvec}[1]{\ensuremath{\bm{\hat{#1}}}}
\newcommand{\Mat}[1]{\ensuremath{\bm{#1}}}

\newcommand{\Mooremat}[2]{\Mat{M}_{#1}( #2 )}

\newcommand{\Gab}[2]{\textrm{Gab}_{#1}\ensuremath{[#2]}}
\newcommand{\GabTransp}[2]{\textrm{Gab}_{#1}^{T}\ensuremath{[#2]}}

\newcommand{\myspace}[1]{\mathcal{#1}}
\newcommand{\Rowspace}[1]{\myspace{R}_q\!\left(#1\right)}

\newcommand{\Colspace}[1]{\myspace{C}_q\left(#1\right)}

\newcommand{\foralldiv}{\ensuremath{\;|\;}}

\newcommand{\swb}{\alpha}
\newcommand{\qpwr}[1]{{[#1]}}

\setlist[itemize]{leftmargin=5mm}
\setlist[enumerate]{leftmargin=5mm}

\usepackage{autonum}

\usepackage[innermargin=.6in,outermargin=.6in,top=.77in,bottom=.77in]{geometry}

 \usepackage{flushend} 

\begin{document}

\begin{acronym}
\acro{RS}[RS]{Reed--Solomon}
\acro{MRD}[MRD]{maximum rank distance}
\end{acronym}

\newcommand{\aw}[1]{{  [\textbf{\textcolor{blue}{#1}} \textcolor{blue!60!black}{--antonia}]}}
\newcommand{\tj}[1]{{  [\textbf{\textcolor{purple}{#1}} \textcolor{purple!60!black}{--thomas}]}}

\title{Decoding of Space-Symmetric Rank Errors
\thanks{This project has received funding from the European Research Council (ERC) under the European Union’s Horizon 2020 research and innovation programme (grant agreement No. 801434)}
}

\author{\IEEEauthorblockN{Thomas Jerkovits}
\IEEEauthorblockA{\textit{Institute of Communication and Navigation} \\
\textit{German Aerospace Center (DLR)}\\
thomas.jerkovits@dlr.de}
\and
\IEEEauthorblockN{Vladimir Sidorenko, Antonia Wachter-Zeh}
\IEEEauthorblockA{\textit{Institute for Communications Engineering} \\
\textit{Technical University of Munich (TUM)}\\
\{vladimir.sidorenko, antonia.wachter-zeh\}@tum.de}
}

\maketitle

\begin{abstract}
This paper investigates the decoding of certain Gabidulin codes that were transmitted over a channel with \emph{space-symmetric errors}. Space-symmetric errors are additive error matrices that have the property that their column and row spaces are equal. We show that for channels restricted to space-symmetric errors, with high probability errors of rank up to $2(n-k)/3$ can be decoded with a Gabidulin code of length $n$ and dimension $k$, using a weak-self orthogonal basis as code locators.
\end{abstract}

\begin{IEEEkeywords}
Gabidulin codes, space-symmetric, rank metric
\end{IEEEkeywords}

\section{Introduction}
Gabidulin codes~\cite{Gabidulin_TheoryOfCodes_1985,Roth_RankCodes_1991,Delsarte_1978} can be considered as the rank-metric analog of Reed--Solomon codes. The rank metric measures the distance between two codewords, represented as matrices, as the rank of their differences. Gabidulin codes are of interest for many applications related to communication, cryptography, space-time coding, network coding, distributed storage systems and digital watermarking~\cite{loidreau2016evolution,loidreau2017new,lusina2003maximum,silva2008rank,silberstein2012error,lefevre2019application}.

Gabidulin codes are \emph{maximum rank distance} (MRD), i.e., their minimum distance is $d_\text{min} = n - k +1$, where $n$ is the length of the code and $k$ the dimension. Hence, it is possible to uniquely decode errors of rank up to $(n-k)/2$. There a several algorithms which efficiently perform unique decoding, e.g.,~\cite{Gabidulin_TheoryOfCodes_1985,Roth_RankCodes_1991,GabidulinParamonovTretjakov_RankErasures_1991,Gabidulin1992Fast,Richter_RankCodes_2004,WachterAfanSido-FastDecGabidulin_DCC_journ}.

In~\cite{Gab05ic,GabidulinPilipchuk_SymmetricRankErrors_2004,GabidulinPilipchuck_SymmMatricesCorrectingErrors_2006} it was shown that for Gabidulin codes that contain a linear subcode of symmetric matrices
 (i.e., the transpose of the matrix coincides with the  matrix itself) can correct symmetric error matrices  of rank  up to $(n -1)/2$.
In this paper, we relax the condition of symmetric errors and consider the case of \emph{space-symmetric error matrices} which have the property that their column and row spaces coincides. We show that it is possible to use a Gabidulin code with the same property as in~\cite{Gab05ic,GabidulinPilipchuk_SymmetricRankErrors_2004,GabidulinPilipchuck_SymmMatricesCorrectingErrors_2006} to decode such space-symmetric errors of rank up to $2(n-k)/3$ with high probability.
We further derive an upper bound on the failure probability of decoding such space-symmetric errors including some simulation results to further support the validation.
Some motivation for the application of space-symmetric errors to code-based cryptography is addressed as well.

\section{Preliminaries}

\subsection{Notation}
Let $q$ be a power of a prime and let $\Fq$ denote the finite field of order $q$ and $\Fqm$ its extension field of order $q^m$.
Denote by $\Fq^{m \times n}$ the set of all $m \times n$ matrices over $\Fq$ and denote the set of all row vectors of length $n$ by $\Fqm^n \defeq \Fqm^{1 \times n}$. For a matrix~$\Mat{A}$, let $A_{i,j}$ be the entry of the $i$-th row and $j$-th column.
For a vector $\vec{\alpha} = (\alpha_0,\alpha_1,\ldots,\alpha_{n-1}) \in \Fqm^n$, define its rank by $\rk(\vec{\alpha}) \defeq \dim \langle \alpha_0,\ldots,\alpha_{n-1} \rangle_q$, where $\langle \alpha_0,\ldots,\alpha_{n-1} \rangle_q$ is the $\Fq$-vector space spanned by the entries $\alpha_i \in \Fqm$.
Given $\alpha \in \Fqm$ and an integer $i$, denote its $i$-th $q$-power by $\alpha^\qpwr{i}$ where $\qpwr{i} = q^i$.
Denote by $\Mooremat{i}{\vec{\alpha}} \in \Fqm^{i \times n}$ the \emph{Moore matrix}
\begin{equation}
\Mooremat{i}{\vec{\alpha}} \defeq \begin{bmatrix}
\swb_0 & \swb_1 & \ldots & \swb_{n-1} \\
\swb_0^\qpwr{1} & \swb_1^\qpwr{1} & \ldots & \swb_{n-1}^\qpwr{1} \\
\vdots &  \ddots      & \vdots & \vdots \\
\swb_0^\qpwr{i-1} & \swb_1^\qpwr{i-1} & \ldots & \swb_{n-1}^\qpwr{i-1} \\
\end{bmatrix}. 
\end{equation}   
We denote the element-wise $j$-th $q$-power of the matrix by $\Mooremat{i}{\vec{\alpha}}^\qpwr{j}$.

Througout this paper, let $m=n$ and $\Mat{A} \in \Fq^{n \times n}$ be a square matrix.
Let $\vec{\alpha} = (\alpha_1,\alpha_2,\ldots,\alpha_n) \in \Fqn^n$  be a fixed basis of $\Fqn$ over $\Fq$. We define the map
\begin{align}
    &\phi : \Fqn^n \to \Fq^{n\times n} \\
    &\vec{a} \mapsto \Mat{A},
\end{align}
where $\vec{a} \in \Fqn^n$ and $\Mat{A} \in \Fq^{n\times n}$ is the unique matrix such that $\vec{a} = \vec{\alpha}\Mat{A}$.
The map $\phi$ is a bijection that preserves the rank and we have that
\begin{equation}
    \rk{(\vec{a})} = \rk{(\Mat{A})}.
\end{equation}
For $\phi(\vec{a})=\Mat{A}$ let $\trvec{a}$ be the vector, such that $\phi(\trvec{a})=\Mat{A}^T$. We call $\trvec{a}$ the \emph{transposed vector} of $\vec{a}$. If $\Mat{A}$ is a symmetric matrix, that means $\Mat{A} = \Mat{A}^T$, then we have that $\vec{a} = \trvec{a}$.

Gabidulin codes are defined by means of \emph{linearized polynomials} which were introduced by Ore \cite{Ore_OnASpecialClassOfPolynomials_1933}.  
A linearized polynomial over $\mathbb{F}_{q^n}$ is a polynomial of the form
$f(x) = \sum_{i=0}^{d_f} f_i x^{[i]}$, 
with $f_i \in \Fqn$. 
If $f_{d_f}\neq 0$, we call $\deg_q f(x) \defeq d_f$ the \textit{q-degree} of $f(x)$. 
An important property of linearized polynomials $\forall \ \alpha_1,\alpha_2 \in \mathbb{F}_{q}$ and $\forall \ a,b \in \mathbb{F}_{q^m}$ is 
$f(\alpha_1 a+\alpha_2 b) = \alpha_1 f(a)+\alpha_2 f(b)$. 
A linearized polynomial of $q$-degree $d$ which contains all elements of a $d$-dimensional subspace as roots is called the \emph{minimal subspace polynomial}.

\subsection{Gabidulin Codes Generated by Weak Self-Orthogonal Bases}
Gabidulin codes \cite{Delsarte_1978,Gabidulin_TheoryOfCodes_1985,Roth_RankCodes_1991} can be seen as the rank-metric analog of \ac{RS} codes and can be defined by a generator matrix as follows.
\begin{definition}[Gabidulin Code]
Denote by $\Gab{\vec{\alpha}}{n,k}$ a Gabidulin code of dimension $k$ and length $n$ over $\Fqn$ which is defined by its $k \times n$ generator matrix
\begin{equation}
\Mat{G}_k \defeq \Mooremat{k}{\vec{\alpha}},
\end{equation}
where $\bm{\alpha} \in \Fqn^n$ and $\alpha_1,\alpha_2,\ldots,\alpha_n$ are linearly independent over $\Fq$.
The set of all Gabidulin codewords is then given by
\begin{equation}
    \Gab{\vec{\alpha}}{n,k} \defeq \{ \vec{u}\Mat{G}_k \foralldiv \forall \vec{u} \in \Fqn^k  \}.
\end{equation}
\end{definition}

Further, we use weak self-orthogonal bases~\cite{MacWilliamsSloane_TheTheoryOfErrorCorrecting_1988,seroussi1980factorization,GabidulinPilipchuck_SymmMatricesCorrectingErrors_2006}.
\begin{definition}[Weak Self-Orthogonal Basis]
A basis $\vec{\alpha} \in \Fqn^{n}$ of $\Fqn$ over $\Fq$ is called a \emph{weak self-orthogonal basis} if
\begin{equation}
    \Mooremat{n}{\vec{\alpha}} \cdot \Mooremat{n}{\vec{\alpha}}^{T} = \Mat{D},
\end{equation}
where $\Mat{D} \in \Fqn^{n \times n}$ is a diagonal matrix.
\end{definition}

\begin{definition}[Transposed Gabidulin Code]
We define the transposed Gabidulin code as
\begin{equation}
    \GabTransp{\vec{\alpha}}{n,k} \defeq \{ \vec{\hat{c}}\foralldiv \forall \vec{c} \in \Gab{\vec{\alpha}}{n,k} \},
\end{equation}
where $\vec{\hat{c}} = \phi^{-1}(\phi(\vec{c})^{T})$.
\end{definition}

If the first row $\vec{\alpha}$ of a generator matrix of a Gabidulin code $\Gab{\vec{\alpha}}{n,k}$ forms a weak self-orthogonal basis, then the parity-check matrix of the code is given by
\begin{equation}
    \Mat{H}_{n-k} = \Mooremat{n-k}{\vec{\alpha}}^\qpwr{k} 
\end{equation}
and the parity-check matrix of the transposed code
$\GabTransp{k}{\vec{\alpha}}$ is given as \cite{GabidulinPilipchuck_SymmMatricesCorrectingErrors_2006} 
\begin{equation}
    \mathbf{\hat{H}}_{n-k} = \Mooremat{n-k}{\vec{\alpha}}^\qpwr{1}. 
\end{equation}

\subsection{Channel Model}
In \cite{GabidulinPilipchuck_SymmMatricesCorrectingErrors_2006,Gab05ic}, the following channel model was considered.
Let the Gabidulin codeword $\vec{c} = \vec{u}\Mat{G}_k$ be corrupted by an error $\vec{e}$ of rank $\rk(\vec{e})=t$, that means
\begin{equation}\label{eq:channel_model}
    \vec{r} = \vec{c} + \vec{e},
\end{equation}
and $\Mat{E} = \phi(\vec{e})$ is a \emph{symmetric} matrix. 
Then, errors of rank up to $t \leq (n-1)/2$ can be corrected for certain parameters \cite{GabidulinPilipchuck_SymmMatricesCorrectingErrors_2006,Gab05ic}.

In this paper, we relax the condition of $\Mat{E}$ being a symmetric matrix, to the condition that the row space of $\Mat{E}$, denoted by $\Rowspace{\Mat{E}}$, equals the column space of $\Mat{E}$, denoted by $\Colspace{\Mat{E}}$, that means
\begin{equation}
    \Rowspace{\Mat{E}} = \Colspace{\Mat{E}}.
\end{equation}
A matrix of rank $t$ whose row space is equal to its column space is called \emph{space-symmetric} and can be decomposed into
\begin{equation}\label{eq:errordecomp}
    \Mat{E} = \Mat{A}\Mat{P}\Mat{A}^{T},
\end{equation}
where $\Mat{A} \in \Fq^{n\times t}$ and $\Mat{P} \in \Fq^{t \times t}$ are full-rank matrices of rank~$t$.
Note that the vector $\vec{a} = (a_0,a_1,\dots,a_{t-1})=\phi^{-1}(\Mat{A})$ is a basis of the column space and also a basis of the row space, since $\Rowspace{\Mat{E}} = \Colspace{\Mat{E}}$.

\section{Decoding Space-Symmetric Errors}
\subsection{Syndrome-Based Decoding Approach}
In the course of this section we introduce a syndrome-based decoding approach (cf.~\cite{Gabidulin_TheoryOfCodes_1985,Roth_RankCodes_1991,GabidulinParamonovTretjakov_RankErasures_1991,Gabidulin1992Fast,Richter_RankCodes_2004} for syndrome-based decoding up to $(n-k)/2$ errors) of Gabidulin codes to decode space-symmetric errors. We therefore show that we can transform the problem of decoding space-symmetric errors into the problem of decoding a special interleaved Gabidulin code of interleaving order two (cf.~\cite{SidorenkoWachterzehChen_InterleavedGab_2012,SB10,Loidreau_Overbeck_Interleaved_2006,WachterzehZeh-InterpolationInterleavedGabidulin_DCC2014} for decoding interleaved Gabidulin codes).
The basic idea is to compute two syndromes, one obtained from the original code and another one by transposing the received noisy codeword matrix and obtaining the syndrome from the transposed Gabidulin code. The two syndromes can then be used to solve a linear system of equations jointly and the decoding radius can be increased beyond $(n-k)/2$. Whether a solution can be found or not, depends on the matrix $\Mat{P}$, see~\eqref{eq:errordecomp}. The explicit decoding approach is similar to decoding a $2$-interleaved Gabidulin code.

From~\eqref{eq:channel_model} we can compute the syndromes
\begin{equation}\label{eq:syndrTr}
    \vec{s}^{(1)} = \vec{\hat{y}}\Mat{\hat{H}}_{n-k}^{T} = \vec{\hat{e}}\Mat{\hat{H}}_{n-k}^{T}
\end{equation}
of the transposed code $\GabTransp{k}{\vec{\alpha}}$ and
\begin{equation}\label{eq:syndrOrg}
    \vec{s}^{(2)} = \vec{y}\Mat{H}_{n-k}^{T} = \vec{e}\Mat{H}_{n-k}^{T}
\end{equation}
of the original $\Gab{k}{\vec{\alpha}}$ code. To each syndrome, we can associate a polynomial $s^{(i)}(x) = \sum_{j=0}^{n-k-1} s_j x^\qpwr{j}$ for $i\in\{1,2\}$.

Given an error decomposed as in~\eqref{eq:errordecomp} we can define the \emph{row error span polynomial} as the minimal subspace polynomial of the vector $\vec{a}$~\cite{Lidl-Niederreiter:FF1996} of degree $t$ as:
\begin{equation}
    \Gamma(x) \defeq \prod_{\vec{u}\in\Rowspace{\Mat{E}}} (x-\phi(\vec{u})).
\end{equation}
Since by definition of the error we have that $\Rowspace{\Mat{E}} = \Colspace{\Mat{E}}$, the row error span polynomial is equal to the column error span polynomial and $\Gamma(a_l) = 0$ for all $l\in\{0,\ldots,t-1\}$.

In the following, we give the key equation of the original code and the transposed code.
\begin{theorem}[Key Equations]
Let $\Gamma(x) = \sum_{i=0}^{t} \Gamma_i x^\qpwr{i}$ be the  error span polynomial with $t = \deg_q \Gamma(x)= \rk{(\vec{e})}$. Then for each syndrome we obtain a key equation as follows
\begin{equation}\label{eq:keyequationorg}
    \Gamma(s^{(i)}(x)) \equiv \Omega^{(i)}(x) \mod x^\qpwr{n-k}, \forall i \in \{1,2\},
\end{equation}
for some $\Omega^{(i)}(x)$ with $\deg_q (\Omega)^{(i)}(x)) < t$. 
\end{theorem}
\begin{IEEEproof}
See Appendix~\ref{sec:appendixke}.
\end{IEEEproof}

Solving the key equation can be done by solving a linear system of equations
\begin{equation}
    \Mat{S}^{(i)}
     \cdot \vec{\Gamma}^{T} = \bm{0},
\end{equation}
where $\vec{\Gamma} = (\Gamma_0,\Gamma_2,\ldots,\Gamma_{t})$ and $\Mat{S}^{(i)}$
\begin{equation}\label{eq:syndromematrix}
    \Mat{S}^{(i)} \defeq \begin{bmatrix}
    {s_{t}^{(i)}}^\qpwr{0} & {s_{t-1}^{(i)}}^\qpwr{1} & \ldots &  {s_{0}^{(i)}}^\qpwr{t} \\
    {s_{t+1}^{(i)}}^\qpwr{0} & {s_{t}^{(i)}}^\qpwr{1} & \ldots &  {s_{1}^{(i)}}^\qpwr{t} \\
    \vdots & \vdots & \ddots & \vdots \\
    {s_{n-k-1}^{(i)}}^\qpwr{0} & {s_{n-k-2}^{(i)}}^\qpwr{1} & \ldots &  {s_{n-k-t}^{(i)}}^\qpwr{t} \\
    \end{bmatrix}.
\end{equation}
Since for each syndrome the error span polynomial in the key equation is the same, we can solve the two key equations jointly.
This approach is similar to decoding a $2$-interleaved Gabidulin code~\cite{SidorenkoWachterzehChen_InterleavedGab_2012,SB10,Loidreau_Overbeck_Interleaved_2006,WachterzehZeh-InterpolationInterleavedGabidulin_DCC2014} which yields the following linear system of equations
\begin{equation}\label{eq:keyequation}
    \Mat{S}\cdot\vec{\Gamma}^{T} = \begin{bmatrix}
    \Mat{S}^{(1)}\\
    \Mat{S}^{(2)}\\
    \end{bmatrix}\cdot\vec{\Gamma}^{T} = \Mat{0},
\end{equation}
where (see Appendix~\ref{sec:appendix})
\begin{equation}\label{eq:syndr1def}
    \Mat{S}^{(1)} = \Mooremat{n-k-t}{\vec{a}}^\qpwr{t+1} \cdot \Mat{P} \cdot \Mooremat{t+1}{\vec{a}}^{T}
\end{equation}
and
\begin{equation}\label{eq:syndr2def}
    \Mat{S}^{(2)} = \Mooremat{n-k-t}{\vec{a}}^\qpwr{t+k} \cdot \Mat{P}^T \cdot \Mooremat{t+1}{\vec{a}}^{T}.
\end{equation}
Thus, $\Mat{S}$ is as follows
\begin{equation}\label{eq:keyequationcomplete}
    \Mat{S} = \begin{bmatrix}
    \Mooremat{n-k-t}{\vec{a}}^\qpwr{t+1} \cdot \Mat{P}\\
    \Mooremat{n-k-t}{\vec{a}}^\qpwr{t+k} \cdot \Mat{P}^T\\
    \end{bmatrix}\cdot \Mooremat{t+1}{\vec{a}}^{T}.
\end{equation}

If $\rk{(\Mat{S})} = t$, we obtain a unique solution for $\Gamma(x)$ up to a scalar factor.
After solving the key equation~\eqref{eq:keyequation} we obtain the coefficients of $\Gamma(x)$ and we can find a basis of the root space of $\Gamma(x)$. This basis corresponds to one possible $\vec{a}$ in the decomposition in~\eqref{eq:errordecomp}. Knowing a possible vector $\vec{a}$, the error can be determined. The complete process of decoding is described in Algorithm~\ref{alg:decodespacesym}. In Appendix~\ref{sec:appendixb}, we describe a way to obtain the error matrix $\Mat{E}$ knowing a possible vector $\vec{a}$. Algorithm~\ref{alg:decodespacesym} has complexity at most $\mathcal{O}(n^3)$ operations over $\Fqn$.

Note that more efficient algorithms with quadratic or even sub-quadratic
complexities in $n$ can be used to solve the joint syndrome key equation from~\eqref{eq:keyequation} as well as to find the matrix $\vec{B}$, see e.g.,~\cite{SidorenkoWachterzehChen_InterleavedGab_2012,SB10,PuchingerWachterzeh-FastOperationsLinearized,PuchingerWachterzeh-ISIT2016}, but for our analysis 
Algorithm~\ref{alg:decodespacesym} is sufficient.

\begin{algorithm}
	\caption{\textsf{DecodeSpaceSymmetric}}\label{alg:decodespacesym}
	\SetKwInOut{Input}{Input}\SetKwInOut{Output}{Output} 
	\Input{$\vec{y}=(y_0,y_1,\ldots,y_n)\in\Fqn^n$,\\Parity-check matrix $\Mat{H}_{n-k}$ of $\Gab{\vec{\alpha}}{n,k}$ }
	\BlankLine
	Syndrome calculations: \\
	$\vec{s}^{(1)} \gets \vec{\hat{y}}\Mat{\hat{H}}_{n-k}^{T} $ and
	$\vec{s}^{(2)} \gets \vec{{y}}\Mat{{H}}_{n-k}^{T} $ \\
    \If{$\vec{s}^{(2)}=\bm{0}$}{
        $\vec{c}\gets\vec{y}$
    }\Else{
    $t\gets \lfloor 2(n-k)/3 \rfloor$\\
    Set up $\Mat{S}^{(1)}$ and $\Mat{S}^{(2)}$ as in~\eqref{eq:syndromematrix}\label{alg:r1}  \\
    $\Mat{S} \gets [(\Mat{S}^{(1)})^T,(\Mat{S}^{(2)})^T]^T$\label{alg:r2}  \\
    \While{$\rk(\Mat{S})<t$}{
    $t\gets t-1$\\
    Repeat~\ref{alg:r1} and~\ref{alg:r2}\\
    Solve: $\Mat{S}\cdot\vec{\Gamma}^T = \bm{0}$ for $\vec{\Gamma} = (\Gamma_0,\ldots,\Gamma_t)\in\Fqn^{t+1}$ \\
    Find a basis $(a_1,a_2,\ldots,a_{\omega}) \in \Fqn^\omega$ of the root space of $\Gamma(x)=\sum_{i=0}^{t}$ \\
    \If{$\omega=t$}{
    Find $\Mat{B}$ such that $\vec{e} = \vec{a}\Mat{B}$ (see Appendix~\ref{sec:appendixb}) \\
    $\vec{c} \gets \vec{y}-\vec{a}\Mat{B}$
    }\Else{
    Declare ``decoding failure''\\
    }
    }
    }
	\BlankLine
	\Output{Estimated codeword $\vec{c}\in\Gab{\vec{\alpha}}{n,k}$ or ``decoding failure''.}

\end{algorithm}

\subsection{Probability of Decoding Failure}
In this section, we show that decoding of space-symmetric errors is guaranteed with high probability.
\begin{theorem}[Decoding of Space-Symmetric Errors]\label{the:maintheorem}
Let $\Gab{\vec{\alpha}}{n,k}$ be given a Gabidulin code of dimension $k$ and length $n$, where  $\vec{\alpha}$ is a weak self-orthogonal basis. Furthermore, let $\vec{r}$ be a noisy Gabidulin codeword as in $\eqref{eq:channel_model}$ where $\Mat{E}$ is a space-symmetric matrix of rank $t \leq 2(n-k)/3$. Then decoding is guaranteed with probability of at least $1-P_f$, where $P_f$ is the decoding failure probability.

Assume that the matrix
\begin{equation}\label{eq:Qmatrix}
 \Mat{Q}\defeq\Mat{P}^{-1}\cdot\Mat{P}^T,
 \end{equation} 
 where $\Mat{P}$ is defined in \eqref{eq:errordecomp}, is uniformly drawn at random from the set of all matrices in $\Fq^{t\times t}$. Then $P_f$ is bounded from above by
\begin{equation}
    P_f \leq 4/q^n.
\end{equation}
\end{theorem}

\begin{IEEEproof}
As discussed above, we obtain a unique solution for $\rk{(\Mat{S})} = t$ to succeed with decoding. To analyze the probability of failure, we restrict to the case for which the matrices $\Mooremat{n-k-t}{\vec{a}}^\qpwr{t+k}$ and $\Mooremat{n-k-t}{\vec{a}}^\qpwr{t+1}$ have no common rows, which means that $t > n-2k$. Consider the case of symmetric error matrices $\Mat{E}$ for which $\Mat{P} = \Mat{P}^T$, we have that 
\begin{equation}
    \Mat{S} = \begin{bmatrix}
    \Mooremat{t+1,n-k+1}{\vec{a}}\\
    \Mooremat{t+k,n}{\vec{a}}\\
    \end{bmatrix} \cdot \Mat{P} \cdot \Mooremat{t+1}{\vec{a}}^{T},
\end{equation}
for which we know that $\rk{(\Mat{P})} = t$ by definition, $\rk{(\Mooremat{t+1}{\vec{a}}^{T})} = t$ and since $n-k < t+k$ also the left part of the decomposition of $\Mat{S}$ has always rank $t$ for $t \leq 2(n-k)/3$.

For the case that $\Mat{P}$ is not symmetric, we can rewrite \eqref{eq:keyequationcomplete} by defining $\Mat{\Tilde{M}}_{n-k-t} \defeq \Mooremat{n-k-t}{\vec{a}}\cdot\Mat{P}$ as
\begin{equation}\label{eq:modS}
    \Mat{S} = \begin{bmatrix}
    \Mat{\Tilde{M}}_{n-k-t}^\qpwr{t+1}\\
    \Mat{\Tilde{M}}_{n-k-t}^\qpwr{t+k} \cdot \Mat{Q}\\
    \end{bmatrix}\cdot \Mooremat{t+1}{\vec{a}}^{T}.
\end{equation}
Assuming that $\Mat{Q}$ is uniformly drawn at random from the set of all matrices in $\Fq^{t\times t}$ 
the matrix $\Mat{S}$ is similar to the syndrome matrix of decoding a $2$-interleaved Gabidulin code and we can bound the probability of decoding error $P_f$ according to~\cite{SB10} and Theorem~\ref{the:maintheorem} follows.
\end{IEEEproof}

\section{Numerical Results}
We simulated a Gabidulin code for $n = 8$, $k = 2$ over $\F_{2^8}$ for a space-symmetric error channel of fixed error weight with $t = \rk(\Mat{E})= 2(n-k)/3 = 4$. The maximum error weight for unique decoding of any rank error is $(n-k)/2 = 3$. We generated $10^6$ noisy Gabidulin codeword samples and we compare the results with a set of different scenarios:
\begin{enumerate}
    \item \label{en:sce1} Space-symmetric errors:
    We draw the matrix $\Mat{A}$ and $\Mat{P}$, both of rank $t$ uniformly at random. Using a Gabidulin code with a weak self-orthonogal basis we decode the nosiy codewords using Algorithm~\ref{alg:decodespacesym}.
     \item \label{en:sce2} Uniform assumption: a modified experiment where we directly draw the matrix $\Mat{Q}$ in~\eqref{eq:Qmatrix}, with $\rk(\Mat{Q})=t$ uniformly at random instead of $\Mat{P}$. We compute the matrix $\Mat{S}$ as in~\eqref{eq:modS} and check its rank. If $\rk(\Mat{S}) \neq t$ we declare a decoding error failure. 
    \item \label{en:sce3} $2$-interleaved Gabidulin code:
    simulation of a $2$-interleaved Gabidulin code where the two error matrices are drawn uniformly at random such that the dimension of its column space is at most $2(n-k)/3 = 4$.
    \item \label{en:sce4} Intersection probability:
    Consider the probability that the intersection of two subspaces $\mathcal{U}$ and $\mathcal{V}$ of $\Fqm^t$  with dimension $\ell$ drawn uniformly at random has dimension larger than or equal to $\omega$
    This probability is~\cite{EV11}
    \begin{equation}\label{eq:estsubintersprob}
    \Pr[\dim{(\mathcal{U} \cap \mathcal{V})} \geq     \omega] =     \frac{\sum_{i=\omega}^{\ell}\binom{t-\ell}{\ell-i}_{q^m}\binom{\ell}{i}_{q^m} \cdot q^{(\ell-i)^2}}{\binom{t}{\ell}_{q^m}}.
    \end{equation}
    Consider the rows of $\Mooremat{n-k-t}{\vec{a}}^\qpwr{t+1} \cdot \Mat{P}$ being a basis of a subspace $\mathcal{\tilde{U}}$ of $\Fqm^t$ of dimension $\ell = n-k-t$. Additionally, consider the rows of $\Mooremat{n-k-t}{\vec{a}}^\qpwr{t+k} \cdot \Mat{P}^T$ being a basis of another subspace $\mathcal{\tilde{V}}$ also of dimension $\ell = n-k-t$. We then can use~\eqref{eq:estsubintersprob} as an estimation of the probability $\Pr[\dim{(\mathcal{\tilde{U}} \cap \mathcal{\tilde{V}})} \geq \omega]$ for $\omega = 2(n-k)-3t+1$ which is equal to the probability of the matrix 
    \begin{equation}
    \begin{bmatrix}
        \Mooremat{t+1,n-k+1}{\vec{a}}\\
        \Mooremat{t+k,n}{\vec{a}}\\        
    \end{bmatrix}
    \end{equation}
    having rank $t$ and therefore $\rk(\Mat{S})=t$ according to~\eqref{eq:keyequationcomplete}. 
\end{enumerate}

Table~\ref{tab:simulationresults} shows the simulation results, including the different scenarios for comparison. We observe that the decoding failure rate of decoding space-symmetric errors using a Gabidulin code with weak self-orthogonal basis is approximately identical to the one with the uniform assumption as well as to the one of decoding a $2$-interleaved Gabidulin code over an ordinary rank-metric channel with errors of a fixed rank. The upper bound on $P_f$ is shown as well and the intersection probability gives a good estimate of the decoding failure rate.

\begin{table}[htbp]
    \centering
    \caption{Simulation results of space-symmetric errors for $n=8$, $k=2$ over $\F_{2^8}$ and $t=4$.}
    \label{tab:simulationresults}
    \begin{tabular}{|l||c|}
	\hline
	Scenario & Decoding failure rate  \\
	\hline\hline
	\ref{en:sce1}) Space-symmetric errors & $0.004124$  \\
	\hline
	\ref{en:sce2}) Uniform assumption &  $0.004229$ \\
	\hline
	\ref{en:sce3}) $2$-interleaved Gabidulin code &  $0.003965$\\
	\hline
	\ref{en:sce4}) Intersection probability & $0.003921$ \\
	\hline
	Upper bound: $4 / q^m$ &  $0.015625$ \\
	\hline
\end{tabular}
\end{table}

\section{Number of Space-Symmetric Matrices}
Denote by $\binom{n}{t}_q$ the Gaussian binomial coefficient which gives the number of $t$-dimensional subspaces of $\Fq^n$ over $\Fq$ and is~\cite{Berlekamp1984Algebraic}
\begin{equation}
    \binom{n}{t}_q = \prod_{i=0}^{t-1} \frac{q^n-q^i}{q^t-q^i}.
\end{equation}
\begin{theorem}[Number of Space-Symmetric Matrices]
The number $\mathcal{N}_\text{sp-sym}(n,t,q)$ of $n \times n$ matrices over $\Fq$ of rank $t$ that are space-symmetric is given by
\begin{equation}\label{eq:nospacesymmatrices}
    \mathcal{N}_\text{sp-sym}(n,t,q)  =  \prod_{i=0}^{t-1} (q^n - q^i).
\end{equation}
\end{theorem}
\begin{IEEEproof}
The number of column spaces of a $n\times n$ matrix of rank $t$ over $\Fq$ is given by the number of $t$-dimensional subspaces of $\Fq^n$ which is $\binom{n}{t}_q$. Since we deal with square matrices we can identify the column space with the image of the associated linear map from $\Fq^n$ to $\Fq^n$. And since column space and row space are equal, there are $\prod_{i=0}^{t-1} (q^t - q^i)$ surjective linear maps from $\Fq^t$ to that $t$-dimensional image. It follows that $\mathcal{N}_\text{sp-sym}(n,t,q) = \binom{n}{t}_q \cdot \prod_{i =0}^{t-1} (q^t - q^i)$ and inserting the definition of $\binom{n}{t}_q$,~\eqref{eq:nospacesymmatrices} follows.
\end{IEEEproof}

\section{Application to Code-Based Cryptography}
A McEliece-like cryptosystem based on Gabidulin codes was first introduced in~\cite{GabidulinParamonovTretjakov_RankErasures_1991}, called the \emph{GPT system}. Unfortunately, the original system and many of its variants were broken by attacks from Gibons~\cite{Gibson1995Severely,Gibson1996Security} and Overbeck~\cite{Overbeck2006Extending,overbeck2005new,overbeck2008structural}. In this section, we present the potential application of space-symmetric rank errors to code-based cryptography. We therefore compare the key sizes of the GPT variant by Loidreau~\cite{loidreau2016evolution,loidreau2017new} if applied to arbitrary rank errors, symmetric errors and space-symmetric errors. We want to emphasize that we do not claim any security proofs. Symmetric errors contain a lot of structure which might lead to new efficient structural attacks when used in cryptosystems like~\cite{loidreau2016evolution,loidreau2017new}. The same holds for space-symmetric errors, however, compared to symmetric errors, the former contain less structure. In either case, for a practical cryptosystem, further analysis to rule out structural attacks is required.

The GPT variant by Loidreau~\cite{loidreau2016evolution,loidreau2017new} involves a parameter~$\lambda$ which amplifies the rank of the error matrix. In Table~\ref{tab:keysizetable1}, we give a set of parameters under the assumption that it is possible to embed error matrices of a specific structure like symmetric or space-symmetric rank errors in the aforementioned cryptosystem. We also give different hypothetical \emph{security levels} (SLs). The SL is defined by the smallest \emph{work factor} (WF) of an attack in bits. 
We assume that the following three WFs (the first two WFs are described in~\cite{loidreau2016evolution}) apply:
\begin{itemize}
    \item Decoding attack: $\text{WF}_\text{dec} = n^3 q^{((t'-1) k}$ 
	\item Structural attack: $\text{WF}_\text{struc} = n^3 q^{n (\lambda-1)-(\lambda-1)^2}$
	\item Brute-forcing error patterns: $\text{WF}_\text{e}$
\end{itemize}
with $t' = t / \lambda$ and $t$ being the maximal amount of errors that can be corrected by the different scenarios:
\begin{enumerate}
    \item Conventional Gabidulin codes: $t = \lfloor (n-k)/2 \rfloor$
    \item Symmetric rank errors: $t = \lfloor (n-1)/2 \rfloor$
    \item Space-symmetric rank errors: $t = \lfloor 2(n-k)/3 \rfloor$
\end{enumerate}
$\text{WF}_\text{e}$ is defined by the number of distinct error matrices which is for the three different cases:
\begin{enumerate}
    \item Conventional rank errors: The number of $n \times n$ matrices of rank $t'$ over $\Fq$ which is given by~\cite{Lidl-Niederreiter:FF1996}
    \begin{equation}
        \mathcal{N}_\text{rank}(n,t',q) = \prod_{j=0}^{t'-1}\frac{(q^n-q^j)^2}{q^{t'}-q^j}.
    \end{equation}
    \item Symmetric rank errors :
    Let $\mathcal{N}_\text{symm}(n,t',q)$ be the number of symmetric matrices of size $n \times n$ of rank $t'=2s$ over $\Fq$ we have that \cite{macwilliams69}
\begin{equation}
\mathcal{N}_\text{symm}(n,2s,q) = \prod_{i=1}^{s}\frac{q^{2i}}{q^{2i}-1} \cdot \prod_{i=0}^{2s-1} (q^{n-i}-1)
\end{equation}
and
\begin{equation}
\mathcal{N}_\text{symm}(n,2s+1,q) = \prod_{i=1}^{s}\frac{q^{2i}}{q^{2i}-1} \cdot \prod_{i=0}^{2s} (q^{n-i}-1).
\end{equation}
    \item Space-symmetric rank errors: $\mathcal{N}_\text{sp-sym}(n,t',q)$ as in~\eqref{eq:nospacesymmatrices}.
\end{enumerate}
Table~\ref{tab:keysizetable1} shows that using symmetric or space-symmetric rank errors potentially might reduce the key size of such a cryptosystem.

\begin{table}
\centering
	\caption{Key sizes of the GPT cryptosystem variant~\cite{loidreau2016evolution,loidreau2017new} using different types of errors: conventional rank errors (Conv), symmetric (Sym) and space-symmetric (Sp-Sym) rank errors for different SLs. The code rate of all codes is approximately $1/2$.}
		\label{tab:keysizetable1}
\begin{tabular}{|c|c|c|c|c|c|c|c|c|c|c|}
	\hline 
	SL & Type & $n$ & $k$ & $\lambda$  & $t'$ & $\text{WF}_\text{dec}$ & $\text{WF}_\text{struc}$ & $\text{WF}_\text{e}$ & Keysize \\ 
	\hline 
	256 & Conv & 96 & 48 & 4  & 6 & 259.75  & 298.75 & 1117.77 & 27.65 KB \\
	256 & Sym & 80 & 40 & 5  & 7 & 258.97  & 322.97 & 539.53 & 16.00 KB \\
	256 & Sp-Sym & 83 & 41 & 4  & 7 & 265.13  & 259.13 & 581.00 & 17.87 KB \\
	\hline
	192 & Conv & 88 & 44 & 4  & 5 & 195.38  & 274.38 & 856.75 & 21.30 KB \\
	192 & Sym & 62 & 31 & 4  & 7 & 203.86  & 194.86 & 413.53 & 7.45 KB \\	
	192 & Sp-Sym & 71 & 35 & 4  & 6 & 193.45  & 222.45 & 426.00 & 11.18 KB \\
	\hline
	128 & Conv & 59 & 29 & 3  & 5 & 133.65  & 131.65 & 566.75 & 6.41 KB \\
	128 & Sym & 49  & 24 & 4  & 6 & 136.84  & 154.84 & 279.53 & 3.68 KB \\
	128 & Sp-Sym & 58  & 29 & 4  & 6 & 162.57  & 129.57 & 348.00 & 6.10 KB \\
	\hline 
\end{tabular} 
\end{table}

\appendix

Define $\Mat{B} \defeq \Mat{P}\Mat{A}^T$ and $\Mat{C} \defeq \Mat{P}^T\Mat{A}^T$. Thus $\Mat{E} = \Mat{A}\Mat{B}$ and $\Mat{E}^T = \Mat{A}\Mat{C}$. The vector representation $\vec{e}$ of $\Mat{E}$ and its transposed $\trvec{e}$ of $\Mat{E}^T$ can therefore be written as
\begin{align}
    \vec{e}=\vec{\alpha}\Mat{E} = \vec{\alpha}\Mat{A}\Mat{B} = \vec{a}\Mat{B}\label{eq:EaB} \\
    \trvec{e}=\vec{\alpha}\Mat{E}^T = \vec{\alpha}\Mat{A}\Mat{C} = \vec{a}\Mat{C},
\end{align}
with $\vec{a} = \vec{\alpha}\Mat{A}$.
From the syndrome equations~\eqref{eq:syndrTr}
 and~\eqref{eq:syndrOrg} follows
 \begin{align}
    \vec{s}^{(1)} = \vec{a}\Mat{C} \hat{\Mat{H}}_{n-k}^{T} \Leftrightarrow s_j^{(1)} &= \sum_{i=0}^{n-1}\sum_{l=0}^{t-1}a_l C_{l,i} \alpha_i^\qpwr{1+j}  \\
    &=\sum_{l=0}^{t-1}a_l \hat{c}_l^\qpwr{1+j}\label{eq:apdecomptr}, \\
    \vec{s}^{(2)} = \vec{a}\Mat{B} {\Mat{H}}_{n-k}^{T} \Leftrightarrow s_j^{(2)} &= \sum_{i=0}^{n-1}\sum_{l=0}^{t-1}a_l B_{l,i} \alpha_i^\qpwr{k+j} \\
    &=\sum_{l=0}^{t-1}a_l \hat{b}_l^\qpwr{k+j}\label{eq:apdecomp},
\end{align}
with $\hat{c}_l$ being the $l$-th entry of the vector $\vec{\hat{c}} = \vec{\alpha}\Mat{C}^T$ and $\hat{b}_l$ of $\vec{\hat{b}} = \vec{\alpha}\Mat{B}^T$, respectively.

\subsection{Proof of the Key Equations}\label{sec:appendixke}
The $p$-th coefficient of $\Omega^{(i)} =\Gamma(s^{(i)}(x))$ for $i\in\{1,2\}$ can be calculated by
\begin{equation}
    \Omega_p^{(i)} = \sum_{j=0}^{p}\Gamma_j (s_{p-j}^{(i)})^\qpwr{j}.
\end{equation}
Using~\eqref{eq:apdecomptr} and~\eqref{eq:apdecomp} we obtain for the transposed code and original code
\begin{equation}
    \Omega_p^{(1)} = \sum_{j=0}^{p} \Gamma_j \left( \sum_{l=0}^{t-1} a_l \hat{c}_l^\qpwr{1+p-j}\right)^\qpwr{j} = \sum_{l=0}^{t-1} \hat{c}_l^\qpwr{1+i}{\sum_{j=0}^{p}\Gamma_j a_l^\qpwr{j}}.
\end{equation}
and
\begin{equation}
    \Omega_p^{(2)} = \sum_{j=0}^{p} \Gamma_j \left( \sum_{l=0}^{t-1} a_l \hat{b}_l^\qpwr{k+p-j}\right)^\qpwr{j} = \sum_{l=0}^{t-1} \hat{b}_l^\qpwr{k+i}{\sum_{j=0}^{p}\Gamma_j a_l^\qpwr{j}}.
\end{equation}
respectively.
For any $p \geq t$ this gives $\Omega_p^{(i)} = 0$, since $\Gamma(a_l)=\sum_{j=0}^{t} \Gamma_j a_l^\qpwr{j}=0$ by definition and therefore $\deg_q \Omega^{(i)}(x) < \deg_q \Gamma(x) = t$ for $i\in\{1,2\}$.

\subsection{Derivation of~\eqref{eq:syndr1def} and~\eqref{eq:syndr2def}}\label{sec:appendix}
Using~\eqref{eq:apdecomptr} and~\eqref{eq:apdecomp} we can decompose the syndrome matrices from~\eqref{eq:syndromematrix} as
\begin{equation}
    \Mat{S}^{(1)} = \begin{bmatrix}
\hat{c}_0^\qpwr{t+1} & \hat{c}_1^\qpwr{t+1} & \ldots & \hat{c}_{t-1}^\qpwr{t+1} \\
\hat{c}_0^\qpwr{t+2} & \hat{c}_1^\qpwr{t+2} & \ldots & \hat{c}_{t-1}^\qpwr{t+2} \\
\vdots & \vdots & \ddots & \vdots \\
\hat{c}_0^\qpwr{n-k} & \hat{c}_1^\qpwr{n-k} & \ldots & \hat{c}_{t-1}^\qpwr{n-k}
\end{bmatrix} \cdot \Mooremat{t+1}{\vec{a}}^{T}
\end{equation}
and
\begin{equation}
\mathbf{S}^{(2)} = \begin{bmatrix}
\hat{b}_0^\qpwr{t+k} & \hat{b}_1^\qpwr{t+k} & \ldots & \hat{b}_{t-1}^\qpwr{t+k} \\
\hat{b}_0^\qpwr{t+k+1} & \hat{b}_1^\qpwr{t+k+1} & \ldots & \hat{b}_{t-1}^\qpwr{t+k+1} \\
\vdots & \vdots & \ddots & \vdots \\
\hat{b}_0^\qpwr{n-1} & \hat{b}_1^\qpwr{n-1} & \ldots & \hat{b}_{t-1}^\qpwr{n-1}
\end{bmatrix} \cdot\Mooremat{t+1}{\vec{a}}^{T}.
\end{equation}
The left sides can be decomposed again according to the definition of $\trvec{c}$ and $\trvec{b}$ and we have
\begin{equation}
\mathbf{S}^{(1)} = \begin{bmatrix}
\swb_0^\qpwr{t+1} & \swb_1^\qpwr{t+1} & \ldots & \swb_{n-1}^\qpwr{t+1} \\
\swb_0^\qpwr{t+2} & \swb_1^\qpwr{t+2} & \ldots & \swb_{n-1}^\qpwr{t+2} \\
\vdots & \vdots & \ddots & \vdots \\
\swb_0^\qpwr{n-k} & \swb_1^\qpwr{n-k} & \ldots & \swb_{n-1}^\qpwr{n-k} 
\end{bmatrix} \cdot \mathbf{C}^\text{T} \cdot \Mooremat{t+1}{\vec{a}}^{T} 
\end{equation}
and
\begin{equation}
\mathbf{S}^{(2)} = \begin{bmatrix}
\swb_0^\qpwr{t+k} & \swb_1^\qpwr{t+k} & \ldots & \swb_{n-1}^\qpwr{t+k} \\
\swb_0^\qpwr{t+k+1} & \swb_1^\qpwr{t+k+1} & \ldots & \swb_{n-1}^\qpwr{t+k+1} \\
\vdots & \vdots & \ddots & \vdots \\
\swb_0^\qpwr{n-1} & \swb_1^\qpwr{n-1} & \ldots & \swb_{n-1}^\qpwr{n-1}
\end{bmatrix} \cdot \mathbf{B}^\text{T} \cdot \Mooremat{t+1}{\vec{a}}^{T} .
\end{equation}
Since $\Mat{C}^T = \Mat{A}\Mat{P}$, $\Mat{B}^T = \Mat{A}\Mat{P}^T$ and $\vec{a} = \vec{\alpha}\Mat{A}$ we obtain~\eqref{eq:syndr1def} and~\eqref{eq:syndr2def}.

\subsection{Finding $\Mat{B}$ such that $\vec{e}=\vec{a}\Mat{B}$}\label{sec:appendixb}
Define $d_l \defeq \hat{b}_l^\qpwr{k}$. We have then from~\eqref{eq:apdecomp} that the syndrome $s_j^{(2)}=\sum_{l=0}^{t-1} a_l d_l^\qpwr{j}$. Knowing the vector $\vec{a} = (a_0,a_1,\ldots,a_{t-1})$ we can solve for $\vec{d} = (d_0,d_1,\ldots,d_{t-1})$ the following linear system of equations:
\begin{equation}
    \begin{pmatrix}
    a_{0}^\qpwr{0} & a_{1}^\qpwr{0} & \cdots & a_{t-1}^\qpwr{0} \\
    a_{0}^\qpwr{-1} & a_{1}^\qpwr{-1} & \cdots & a_{t-1}^\qpwr{-1} \\
    \vdots & \vdots & \ddots & \vdots \\
    a_{0}^\qpwr{-v} & a_{1}^\qpwr{-v} & \cdots & a_{t-1}^\qpwr{-v} \\
    \end{pmatrix}\cdot \begin{pmatrix}
    d_0 \\
    d_1 \\
    \vdots \\
    d_{t-1}  
    \end{pmatrix} = \begin{pmatrix}
    (s_0^{(2)})^\qpwr{0} \\
    (s_1^{(2)})^\qpwr{-1} \\
    \vdots \\
    (s_v^{(2)})^\qpwr{-v} \\
    \end{pmatrix}
\end{equation}
with $v = n-k-1$. It remains to find $\Mat{B}$ such that $d_l = \sum_{j=0}^{n-1} B_{l,j} \alpha_j^\qpwr{k}$.

\bibliographystyle{IEEEtran}

\begin{thebibliography}{10}
\providecommand{\url}[1]{#1}
\csname url@samestyle\endcsname
\providecommand{\newblock}{\relax}
\providecommand{\bibinfo}[2]{#2}
\providecommand{\BIBentrySTDinterwordspacing}{\spaceskip=0pt\relax}
\providecommand{\BIBentryALTinterwordstretchfactor}{4}
\providecommand{\BIBentryALTinterwordspacing}{\spaceskip=\fontdimen2\font plus
\BIBentryALTinterwordstretchfactor\fontdimen3\font minus
  \fontdimen4\font\relax}
\providecommand{\BIBforeignlanguage}[2]{{%
\expandafter\ifx\csname l@#1\endcsname\relax
\typeout{** WARNING: IEEEtran.bst: No hyphenation pattern has been}%
\typeout{** loaded for the language `#1'. Using the pattern for}%
\typeout{** the default language instead.}%
\else
\language=\csname l@#1\endcsname
\fi
#2}}
\providecommand{\BIBdecl}{\relax}
\BIBdecl

\bibitem{Gabidulin_TheoryOfCodes_1985}
E.~M. Gabidulin, ``{Theory of Codes with Maximum Rank Distance},'' \emph{Probl.
  Inf. Transm.}, vol.~21, no.~1, pp. 3--16, 1985.

\bibitem{Roth_RankCodes_1991}
R.~M. Roth, ``{Maximum-Rank Array Codes and their Application to Crisscross
  Error Correction},'' \emph{IEEE Trans. Inf. Theory}, vol.~37, no.~2, pp.
  328--336, 1991.

\bibitem{Delsarte_1978}
P.~Delsarte, ``{Bilinear Forms over a Finite Field with Applications to Coding
  Theory},'' \emph{J. Combin. Theory}, vol.~25, no.~3, pp. 226--241, 1978.

\bibitem{loidreau2016evolution}
P.~Loidreau, ``{An Evolution of GPT Cryptosystem}.''\hskip 1em plus 0.5em minus
  0.4em\relax ACCT, 2016.

\bibitem{loidreau2017new}
------, ``A new rank metric codes based encryption scheme,'' in
  \emph{International Workshop on Post-Quantum Cryptography}.\hskip 1em plus
  0.5em minus 0.4em\relax Springer, 2017, pp. 3--17.

\bibitem{lusina2003maximum}
P.~Lusina, E.~M. Gabidulin, and M.~Bossert, ``{Maximum Rank Distance Codes as
  Space-Time Codes},'' \emph{IEEE Trans. Inform. Theory}, vol.~49, no.~10, pp.
  2757--2760, Oct. 2003.

\bibitem{silva2008rank}
D.~Silva, F.~R. Kschischang, and R.~Koetter, ``A rank-metric approach to error
  control in random network coding,'' \emph{IEEE transactions on information
  theory}, vol.~54, no.~9, pp. 3951--3967, 2008.

\bibitem{silberstein2012error}
N.~Silberstein, A.~S. Rawat, and S.~Vishwanath, ``Error resilience in
  distributed storage via rank-metric codes,'' in \emph{2012 50th Annual
  Allerton Conference on Communication, Control, and Computing
  (Allerton)}.\hskip 1em plus 0.5em minus 0.4em\relax IEEE, 2012, pp.
  1150--1157.

\bibitem{lefevre2019application}
P.~Lef{\`e}vre, P.~Carr{\'e}, and P.~Gaborit, ``Application of rank metric
  codes in digital image watermarking,'' \emph{Signal Processing: Image
  Communication}, vol.~74, pp. 119--128, 2019.

\bibitem{GabidulinParamonovTretjakov_RankErasures_1991}
E.~M. Gabidulin, A.~V. Paramonov, and O.~V. Tretjakov, ``{Rank Errors and Rank
  Erasures Correction},'' in \emph{Int. Colloq. Coding Theory}, 1991.

\bibitem{Gabidulin1992Fast}
E.~M. Gabidulin, ``{A Fast Matrix Decoding Algorithm for Rank-Error-Correcting
  Codes},'' \emph{Algebraic Coding}, vol. 573, pp. 126--133, 1992.

\bibitem{Richter_RankCodes_2004}
G.~Richter and S.~Plass, ``{Fast Decoding of Rank-Codes with Rank Errors and
  Column Erasures},'' in \emph{IEEE Int. Symp. Inf. Theory (ISIT)}, 2004, p.
  398.

\bibitem{WachterAfanSido-FastDecGabidulin_DCC_journ}
A.~{Wachter-Zeh}, V.~Afanassiev, and V.~Sidorenko, ``{Fast Decoding of
  {G}abidulin Codes},'' \emph{Des. Codes Cryptogr.}, vol.~66, no.~1, pp.
  57--73, Jan. 2013.

\bibitem{Gab05ic}
N.~Pilipchuk and E.~Gabidulin, ``On codes correcting symmetric rank errors,''
  vol. 3969, 01 2005, pp. 14--21.

\bibitem{GabidulinPilipchuk_SymmetricRankErrors_2004}
E.~M. Gabidulin and N.~I. Pilipchuk, ``{Symmetric Rank Codes},'' \emph{Probl.
  Inf. Transm.}, vol.~40, no.~2, pp. 103--117, 2004.

\bibitem{GabidulinPilipchuck_SymmMatricesCorrectingErrors_2006}
------, ``{Symmetric matrices and codes correcting rank errors beyond the
  [(d-1)/2] bound},'' \emph{Discrete Applied Mathematics}, vol. 154, no.~2, pp.
  305--312, 2006.

\bibitem{Ore_OnASpecialClassOfPolynomials_1933}
{\O}.~Ore, ``{On a Special Class of Polynomials},'' \emph{Trans. Amer. Math.
  Soc.}, vol.~35, pp. 559--584, 1933.

\bibitem{MacWilliamsSloane_TheTheoryOfErrorCorrecting_1988}
F.~J. MacWilliams and N.~J.~A. Sloane, \emph{{The Theory of Error-Correcting
  Codes}}.\hskip 1em plus 0.5em minus 0.4em\relax North Holland Publishing Co.,
  1988.

\bibitem{seroussi1980factorization}
G.~Seroussi and A.~Lempel, ``Factorization of symmetric matrices and
  trace-orthogonal bases in finite fields,'' \emph{SIAM Journal on Computing},
  vol.~9, no.~4, pp. 758--767, 1980.

\bibitem{SidorenkoWachterzehChen_InterleavedGab_2012}
V.~R. Sidorenko, A.~{Wachter-Zeh}, and D.~Chen, ``On fast {D}ecoding of
  {I}nterleaved {G}abidulin {C}odes,'' in \emph{{I}nt. {S}ymp. {P}robl.
  {R}edundancy {I}nf. {C}ontrol {S}ystems}, Sep. 2012, pp. 78--83.

\bibitem{SB10}
V.~{Sidorenko} and M.~{Bossert}, ``Decoding interleaved {G}abidulin codes and
  multisequence linearized shift-register synthesis,'' in \emph{2010 IEEE
  International Symposium on Information Theory}, 2010, pp. 1148--1152.

\bibitem{Loidreau_Overbeck_Interleaved_2006}
P.~Loidreau and R.~Overbeck, ``{Decoding Rank Errors Beyond the Error
  Correcting Capability},'' in \emph{Int. Workshop Alg. Combin. Coding Theory
  (ACCT)}, Sep. 2006, pp. 186--190.

\bibitem{WachterzehZeh-InterpolationInterleavedGabidulin_DCC2014}
A.~Wachter-Zeh and A.~Zeh, ``{List and Unique Error-Erasure Decoding of
  Interleaved Gabidulin Codes with Interpolation Techniques},'' \emph{Des.
  Codes Cryptogr.}, vol.~73, no.~2, pp. 547--570, 2014.

\bibitem{Lidl-Niederreiter:FF1996}
R.~Lidl and H.~Niederreiter, \emph{{Finite Fields}}, ser. Encyclopedia of
  Mathematics and its Applications.\hskip 1em plus 0.5em minus 0.4em\relax
  Cambridge University Press, Oct. 1996.

\bibitem{PuchingerWachterzeh-FastOperationsLinearized}
S.~Puchinger and A.~{Wachter-Zeh}, ``{Fast Operations on Linearized Polynomials
  and their Applications in Coding Theory},'' \emph{Journal of Symbolic
  Computation}, vol.~89, pp. 194--215, Nov. 2018.

\bibitem{PuchingerWachterzeh-ISIT2016}
------, ``{Sub-quadratic Decoding of Gabidulin Codes},'' in \emph{IEEE Int.
  Symp. Inf. Theory (ISIT)}, Jul. 2016.

\bibitem{EV11}
T.~{Etzion} and A.~{Vardy}, ``Error-correcting codes in projective space,''
  \emph{IEEE Transactions on Information Theory}, vol.~57, no.~2, pp.
  1165--1173, 2011.

\bibitem{Berlekamp1984Algebraic}
E.~R. Berlekamp, \emph{{Algebraic Coding Theory}}, revised~ed.\hskip 1em plus
  0.5em minus 0.4em\relax Aegean Park Press, Jun. 1984.

\bibitem{Gibson1995Severely}
\BIBentryALTinterwordspacing
J.~K. Gibson, ``{Severely denting the Gabidulin version of the McEliece Public
  Key Cryptosystem},'' \emph{Des. Codes Cryptogr.}, vol.~6, no.~1, pp. 37--45,
  Jul. 1995. [Online]. Available: \url{http://dx.doi.org/10.1007/BF01390769}
\BIBentrySTDinterwordspacing

\bibitem{Gibson1996Security}
K.~Gibson, ``{The Security of the Gabidulin Public Key Cryptosystem},''
  \emph{Advances in Cryptology}, vol. 1070, pp. 212--223, 1996.

\bibitem{Overbeck2006Extending}
R.~Overbeck, ``{Extending Gibson's Attacks on the GPT Cryptosystem},''
  \emph{Coding and Cryptography --- Revised selected papers of WCC 2005}, vol.
  3969, pp. 178--188, 2006.

\bibitem{overbeck2005new}
------, ``A new structural attack for gpt and variants,'' in
  \emph{International Conference on Cryptology in Malaysia}.\hskip 1em plus
  0.5em minus 0.4em\relax Springer, 2005, pp. 50--63.

\bibitem{overbeck2008structural}
------, ``Structural attacks for public key cryptosystems based on gabidulin
  codes,'' \emph{Journal of cryptology}, vol.~21, no.~2, pp. 280--301, 2008.

\bibitem{macwilliams69}
\BIBentryALTinterwordspacing
J.~MacWilliams, ``Orthogonal matrices over finite fields,'' \emph{The American
  Mathematical Monthly}, vol.~76, no.~2, pp. 152--164, 1969. [Online].
  Available: \url{http://www.jstor.org/stable/2317262}
\BIBentrySTDinterwordspacing

\end{thebibliography}

\end{document}